\pdfoutput=1

\documentclass[twocolumn,aps,prl,superscriptaddress,showpacs]{revtex4}

\usepackage{times}
\usepackage{graphicx}
\usepackage{epsfig}
\usepackage{amsfonts}
\usepackage{amsmath}
\usepackage{amssymb}
\usepackage{color}
\usepackage{multirow}
\usepackage[colorlinks=true,citecolor=blue,urlcolor=blue]{hyperref}

\def\endproof{\vrule height6pt width6pt depth0pt} 

\begin{document}


\title{Simple Explanation of the Quantum Limits of Genuine $n$-Body Nonlocality}


\author{Ad\'an~Cabello}
 \affiliation{Departamento de F\'{\i}sica Aplicada II, Universidad de
 Sevilla, E-41012 Sevilla, Spain}


\date{\today}


\begin{abstract}
Quantum $n$-body correlations cannot be explained with $(n-1)$-body nonlocality. However, this genuine $n$-body nonlocality cannot surpass certain bounds. Here we address the problem of identifying the principles responsible for these bounds. We show that, for any $n \ge 2$, the exclusivity principle, as derived from axioms about sharp measurements, and a technical assumption give the exact bounds predicted by quantum theory. This provides a unified explanation of the bounds of single-body contextuality and $n$-body nonlocality, and connects two programs towards understanding quantum theory.
\end{abstract}



\maketitle


{\em Introduction.---}Bell showed that quantum theory (QT) violates bipartite inequalities satisfied by local realistic theories, called Bell inequalities \cite{Bell64}. Similar violations occur for systems of three or more parties \cite{Mermin90}. However, this so-called ``quantum nonlocality'' is just the tip of an iceberg. Svetlichny showed that QT also violates tripartite inequalities satisfied by theories in which arbitrary correlations (even signaling ones) are allowed between two of the three parties, but only local correlations are allowed between these two parties and the third one; the pair with arbitrary correlations may be different in each repetition of the experiment \cite{Svetlichny87}. Indeed, as shown independently by Collins {\em et al.}\ \cite{CGPRS02} and Seevinck and Svetlichny \cite{SS02}, this phenomenon is even more general: QT violates $n$-partite inequalities satisfied by theories in which arbitrary correlations are allowed among $k$ parties (with $1 \le k \le n-1$) and also among the other $n-k$ parties, while only local correlations are allowed between the first $k$ parties and the second $n-k$ parties; again, the sets of $k$ and $n-k$ parties may be different in each repetition of the experiment. This means that $(n-1)$-partite nonlocality cannot explain $n$-partite quantum correlations. In this sense, QT exhibits ``genuine $n$-body nonlocality.'' Recently, genuine three-body nonlocality has been observed in experiments with photons \cite{LKR09}, even under strict locality conditions \cite{EMFLHYPBPSRGLWJR14}, and also with high detection efficiency \cite{HSHMMVMNRJ14}. In addition, genuine multipartite quantum nonlocality free of the detection loophole has been observed with trapped ions for up to $n=6$ \cite{BBSNHMGB13}.

The existence of a gap between the maximum quantum violation of the inequalities detecting genuine multipartite nonlocality and the maximum violation allowed by nonsignaling \cite{Svetlichny87, CGPRS02, SS02} adds a new perspective to Popescu and Rohrlich's question of why nature is exactly as nonlocal as predicted by QT, but not more or less nonlocal \cite{PR94}. The question now becomes why is nature exactly as genuinely $n$-body nonlocal.

There are two different programs towards understanding quantum correlations from first principles. On one hand, there is the ``black box'' program, which deals with correlations produced by local demolition measurements in multipartite scenarios. This approach makes no assumptions on the internal working of the measurement devices (which are treated as black boxes) and pays no attention to the postmeasurement state and, consequently, restricts its attention to a particular class of ``quantum correlations,'' namely those represented by joint probabilities $p(a_1,\ldots,a_n|x_1,\ldots,x_n)$ of obtaining outcome $a_i$ when measurement $x_i$ is performed on body $X_i$, assuming that the recording of $a_i$ and the choice of $x_j$ can be spacelike separated for all pairs in which $i\neq j$. Although this approach has explained some quantum correlations \cite{PPKSWZ09}, including the quantum limits of the inequalities detecting genuine $n$-body nonlocality \cite{XR11}, it cannot explain quantum correlations in contextuality scenarios \cite{Cabello11}. Moreover, it has been proven that the principle used in Refs.~\cite{PPKSWZ09,XR11} cannot explain the impossibility of some nonquantum extremal nonlocal boxes in the three-party, two-setting, two-outcome scenario \cite{YCATS12}. In addition, there is increasing evidence that the black box program, with independence of the principle invoked, cannot explain all nonlocal quantum correlations \cite{NGHA14}.

In contrast, the ``sharp measurements,'' ``graph-theoretic,'' or ``contextuality'' program \cite{CSW10, CSW14, RDLTC14} deals with correlations produced by ``sharp'' measurements, as defined in the framework of general probabilistic theories (GPTs) in Ref.~\cite{CY14}, namely, a measurement is sharp when it is repeatable (i.e., it does not perturb any subsequent implementation of the same measurement) and minimally disturbing (i.e., it does not disturb the other measurements with which it is compatible); see also Ref.~\cite{Kleinmann14}. Within this program, ``quantum correlations'' are those represented by joint probabilities $p(a,\ldots,c|x,\ldots,z)$ in which $x,\ldots,z$ are sharp and compatible (but not necessarily spacelike separated) measurements. For sharp measurements, an ``event'' is a transformation that turns the initial state of the system into a new state. An event is characterized by the outcomes of a set of sharp measurements on the initial state; e.g., an event can be labeled as $(a,\ldots,c|x,\ldots,z)$. Two events are equivalent if they correspond to indistinguishable transformations. Two events are exclusive if there exists a sharp measurement that contains both of them. Within this program, the exclusivity (E) principle \cite{Cabello13} states that any set of $m$ pairwise exclusive events is $m$-wise exclusive. Therefore, from Kolmogorov's axioms of probability, the sum of the probabilities of $m$ pairwise exclusive events cannot exceed~1. However, the E principle is not implied by Kolmogorov's axioms; there are theories satisfying Kolmogorov's axioms which do not satisfy the E principle \cite{Specker60}. On the other hand, the E principle is implied by some fundamental principles proposed to derive QT \cite{CY14, BMU14, Henson14}.

The E principle has explained the maximum quantum violation of the simplest noncontextuality inequality \cite{Cabello13} and excluded any set of probability distributions strictly larger than the set of correlations achievable with quantum systems for any self-complementary exclusivity graph \cite{ATC14}. The exclusivity graph of a set of events is the graph in which each event is represented by a vertex and there is an edge if and only if the events at the end points are exclusive. By making an additional assumption, the E principle has also explained the Tsirelson bound of the Clauser-Horne-Shimony-Holt (CHSH) Bell inequality \cite{Cabello14}. The E principle applied to the device-independent program is called local orthogonality and has explained why all extremal nonlocal boxes in the three-party, two-setting, two-outcome scenario are not feasible in QT \cite{FSABCLA13}.

The aim of this Letter is to show that, within the framework of GPTs as defined in, e.g., Ref.~\cite{CY14}, the limits of the $n$-body nonlocality predicted by QT for the scenario considered in Refs.~\cite{CGPRS02, SS02}, namely, $n$ parties, each with two possible measurements, and each with two possible outcomes, are exactly singled out by the following assumptions: (i) Every measurement can be realized as a sharp measurement on the system and the environment. (ii) A sharp measurement on system $A$ and a sharp measurement on system $A'$ is a sharp measurement on the composite system $AA'$. (iii) The coarse graining of a sharp measurement is a sharp measurement. (iv) There exist some dichotomic measurements $x_1=0$ and $x_1=1$ on system $A$, and $x'_1=0$ and $x'_1=1$ on system $A'$ such that $A_{00}$ and $A_{11}$ are compatible and $A_{01}$ and $A_{10}$ are compatible, where $A_{ij}$ are dichotomic measurements on $AA'$ giving $0$ if the results of $x_1=i$ and $x'_1=j$ are equal, and $1$ otherwise. Assumptions (ii) and (iii) imply the E principle \cite{CY14}.


{\em Maximum quantum violation of inequalities to detect genuine $n$-body nonlocality.---}We focus on the inequalities to detect genuine $n$-body nonlocality, with $n \ge 2$, introduced, independently, by Collins {\em et al.}\ \cite{CGPRS02} and Seevinck and Svetlichny \cite{SS02}. There is one inequality for each $n$. It can be written, together with its maximum quantum violation, as
\begin{equation}
 S_n \stackrel{\mbox{\tiny{H}}}{\leq} 2^{n-1} \stackrel{\mbox{\tiny{QT}}}{\leq} \sqrt{2} \times 2^{n-1},
 \label{ineq}
\end{equation}
where ``$\stackrel{\mbox{\tiny{H}}}{\leq} 2^{n-1}$'' indicates the tight bound for hybrid theories in which arbitrary correlations are allowed among $k$ parties (with $1\le k \le n-1$) and also among the other $n-k$ parties, while only local correlations are allowed between the first $k$ parties and the second $n-k$ parties; and ``$\stackrel{\mbox{\tiny{QT}}}{\leq} \sqrt{2} \times 2^{n-1}$'' indicates the tight bound in QT. $S_n$ can be defined recursively from
\begin{equation}
 S_2 = \langle x_0 x_0 \rangle + \langle x_0 x_1 \rangle + \langle x_1 x_0 \rangle - \langle x_1 x_1 \rangle,
\end{equation}
where $\langle x_0 x_1 \rangle$ denotes the mean value of the product of the outcomes ($+1$ or $-1$) of $x_0$, measured by the first party, and $x_1$, measured by the second party. The recursive formula is
\begin{equation}
 S_n = \langle S_{n-1} x_1 \rangle + \langle \bar{S}_{n-1} x_0 \rangle,
\end{equation}
where $\bar{S}_{n-1}$ is obtained from $S_{n-1}$ by exchanging $x_0$ and $x_1$, and $\langle S_{n-1} x_1 \rangle$ must be interpreted as in the following example:
\begin{equation}
 \begin{split}
 S_3 = & \langle x_0 x_0 x_1 \rangle + \langle x_0 x_1 x_1 \rangle + \langle x_1 x_0 x_1 \rangle - \langle x_1 x_1 x_1 \rangle \\
 & + \langle x_1 x_1 x_0 \rangle + \langle x_1 x_0 x_0 \rangle + \langle x_0 x_1 x_0 \rangle - \langle x_0 x_0 x_0 \rangle.
 \end{split}
\end{equation}
For $n=2$, the inequality reduces to the CHSH inequality and the quantum bound to the Tsirelson bound.


{\em Explanation of the quantum bounds.---}For convenience, we relabel party $i$'s measurement outcomes $a_i \in \{+1,-1\}$ as $b_i=(1 - a_i)/2$, so that $b_i \in \{0,1\}$. Then, expression (\ref{ineq}) can be rewritten as
\begin{equation}
 \Sigma_n \stackrel{\mbox{\tiny{H}}}{\leq} 3 \times 2^{n-2} \stackrel{\mbox{\tiny{QT}}}{\leq} (2 + \sqrt{2}) \times 2^{n-2},
 \label{newineq}
\end{equation}
with
\begin{align}
\label{sigma}
 \Sigma_n = &\sum_{
\begin{array}{c}
(x_1,x_2,x_3)\neq(x_1,x_1,x_1)\\
b_1 \oplus \cdots \oplus b_n =0\\
\end{array}}
 p(b_1,\ldots,b_n|x_1,\ldots,x_n) \nonumber \\ &+
 \sum_{
\begin{array}{c}
(x_1,x_2,x_3)=(x_1,x_1,x_1)\\
b_1 \oplus \cdots \oplus b_n =1\\
\end{array}}
 p(b_1,\ldots,b_n|x_1,\ldots,x_n),
\end{align}
where the first sum extends to all $x_1,\ldots,x_n$ such that $x_1$, $x_2$, and $x_3$ are not all equal, and to all $b_1,\ldots,b_n$ such that $b_1 \oplus \cdots \oplus b_n =0$, where $\oplus$ denotes addition modulo~2. Similarly, the second sum extends to all $x_1,\ldots,x_n$ such that $x_1=x_2=x_3$, and to all $b_1,\ldots,b_n$ such that $b_1 \oplus \cdots \oplus b_n =1$. The relation between $S_n$ and $\Sigma_n$ is
\begin{equation}
 \Sigma_n = \frac{S_n}{2} + 2^{n-1}.
\end{equation}

In the framework of GPTs, it is assumed that the state of two systems can be prepared and measured independently, and that the same experiment can be done twice, generating exactly the same probability distribution.
Let us consider two independent copies of an experiment testing $\Sigma_n$ for a given $n\ge2$; one copy in Stockholm and the other in Vienna. If we denote an event in Stockholm by $(b_1,\ldots,b_n|x_1,\ldots,x_n)$ and an event in Vienna by $(b'_1,\ldots,b'_n|x'_1,\ldots,x'_n)$, the probability of the global event $(b_1,\ldots,b_n,b'_1,\ldots,b'_n|x_1,\ldots,x_n,x'_1,\ldots,x'_n)$ is
\begin{align}
 &p(b_1,\ldots,b_n,b'_1,\ldots,b'_n|x_1,\ldots,x_n,x'_1,\ldots,x'_n) \nonumber \\
 &=p(b_1,\ldots,b_n|x_1,\ldots,x_n) \times p(b'_1,\ldots,b'_n|x'_1,\ldots,x'_n).
 \label{indep}
\end{align}

The sum of the probabilities of the events $(b_1,\ldots,b_n,b'_1,\ldots,b'_n|x_1,\ldots,x_n,x'_1,\ldots,x'_n)$ in which both the Stockholm and Vienna parts are in $\Sigma_n$ is $\Sigma_n^2$. Similarly, the sum of the probabilities of the events of the type $(b_1,\ldots,b_n,b'_1,\ldots,b'_n|x_1,\ldots,x_n,x'_1,\ldots,x'_n)$ in which both the Stockholm and Vienna parts are not in $\Sigma_n$ is $(2^n-\Sigma_n)^2$. The term $2^n$ comes from the fact that there are $2^n$ different ``contexts'' $(x_1,\ldots,x_n)$ in $\Sigma_n$ and that the sum of the probabilities of the events in a context is~1.

In these two sums there are, in total, $(4^n \times 4^n)/2$ events $(b_1,\ldots,b_n,b'_1,\ldots,b'_n|x_1,\ldots,x_n,x'_1,\ldots,x'_n)$. These events can be distributed in $4^n$ disjoint sets, each set containing $2 \times 4^{n-1}$ pairwise exclusive events. These disjoint sets are constructed recursively from the corresponding sets for $n=2$, which are described in Table~\ref{Table2}.


\begin{table*}[tb]
\begin{tabular}{lcr}
\begin{tabular}{ccccccc}
\hline
\hline
\cline{1-7}
$(0,0,0,0|0,0,0,0)$& &$(0,1,0,1|0,0,0,0)$& &$(0,0,1,1|1,0,1,0)$& &$(0,1,1,0|1,0,1,0)$\\
$(0,0,1,1|0,0,0,0)$& &$(0,1,1,0|0,0,0,0)$& &$(0,0,0,0|1,0,1,0)$& &$(0,1,0,1|1,0,1,0)$\\
$(0,0,1,1|0,0,0,1)$& &$(0,1,1,0|0,0,0,1)$& &$(0,0,1,0|1,0,1,1)$& &$(0,1,1,1|1,0,1,1)$\\
$(0,0,0,0|0,0,0,1)$& &$(0,1,0,1|0,0,0,1)$& &$(0,0,0,1|1,0,1,1)$& &$(0,1,0,0|1,0,1,1)$\\
$(0,0,1,0|0,0,1,1)$& &$(0,1,1,1|0,0,1,1)$& &$(0,0,1,1|1,0,0,1)$& &$(0,1,1,0|1,0,0,1)$\\
$(0,0,0,1|0,0,1,1)$& &$(0,1,0,0|0,0,1,1)$& &$(0,0,0,0|1,0,0,1)$& &$(0,1,0,1|1,0,0,1)$\\
$(0,0,0,0|0,0,1,0)$& &$(0,1,0,1|0,0,1,0)$& &$(0,0,1,1|1,0,0,0)$& &$(0,1,1,0|1,0,0,0)$\\
$(0,0,1,1|0,0,1,0)$& &$(0,1,1,0|0,0,1,0)$& &$(0,0,0,0|1,0,0,0)$& &$(0,1,0,1|1,0,0,0)$\\
\cline{1-7}
\hline
\hline
\end{tabular}
& \;\;\;\; &
\begin{tabular}{c}
\hline
\hline
\cline{1-1}
 $(1,0|A_{00},A_{11})$\\
 $(0,1|A_{00},A_{11})$\\
 $(0,0|A_{00},A_{11})$\\
 $(1,1|A_{00},A_{11})$\\
 $(0,0|A_{01},A_{10})$\\
 $(1,1|A_{01},A_{10})$\\
 $(1,0|A_{01},A_{10})$\\
 $(0,1|A_{01},A_{10})$\\
\cline{1-1}
\hline
\hline
\end{tabular}
\\
\end{tabular}
\caption{Left-hand side (i.e., first four columns): Each row represents a set of eight pairwise exclusive events of the type $(b_1,b_2,b'_1,b'_2|x_1,x_2,x'_1,x'_2)$.
Only four events are displayed; the other four are obtained by applying to each of the four events the transformation $(b_1,b_2,b'_1,b'_2|x_1,x_2,x'_1,x'_2) \rightarrow (b_1 \oplus 1,b_2 \oplus 1,b'_1 \oplus 1,b'_2 \oplus 1|x_1,x_2,x'_1,x'_2)$, where $\oplus$ is the sum modulo $2$. Other eight sets of eight pairwise exclusive events can be obtained by exchanging the roles of the cities, i.e., by applying to each of the elements of each of the previous eight sets the transformation $(b_1,b_2,b'_1,b'_2|x_1,x_2,x'_1,x'_2) \rightarrow (b'_1,b'_2,b_1,b_2|x'_1,x'_2,x_1,x_2)$. The resulting 16 sets of eight pairwise exclusive events are disjoint.
Right-hand side (i.e., fifth column): Event of the type $(c,d|A_{00},A_{11})$ or $(e,f|A_{01},A_{10})$ which is exclusive to each of the eight events in the same row of the left-hand side. The corresponding event for each of the other eight sets not displayed in the table is obtained by applying to the corresponding event in the right-hand side the transformation $(g,h|A_{ii'},A_{jj'}) \rightarrow (g,h|A_{i'i},A_{j'j})$.
\label{Table2}}
\end{table*}


Given the sets of pairwise exclusive events $(b_1,\ldots,b_n,b'_1,\ldots,b'_n|x_1,\ldots,x_n,x'_1,\ldots,x'_n)$ for $n$ parties, the corresponding sets of pairwise exclusive events for $n+1$ parties are constructed as follows.
Take each one of the sets for $n$ parties and add, to each of the events, new components $(b_{n+1},b'_{n+1}|x_{n+1},x'_{n+1})$, with $(x_{n+1},x'_{n+1})$ chosen from $\{(0,0),(0,1),(1,0),(1,1)\}$ and all $(b_{n+1},b'_{n+1})\in\{(0,0),(0,1),(1,0),(1,1)\}$. This way, each set for $n$ parties gives rise to four sets for $n+1$ parties; each of them having four times more elements than the original set for $n$ parties. Since the sets for $n=2$ parties are disjoint, the sets for arbitrary $n$ are also disjoint.

Now recall that we are assuming that measurements are sharp. Therefore, there is the possibility of performing extra measurements after those used to label the events $(b_1,\ldots,b_n,b'_1,\ldots,b'_n|x_1,\ldots,x_2,x'_1,\ldots,x'_2)$. Let us consider four possible measurements $A_{ij}$, with $i,j \in \{0,1\}$. $A_{ij}$ is a measurement on the particle of the first party in Stockholm and the particle of the first party in Vienna. By definition, $A_{ij}$ gives $0$ if $b_1=b'_1$ (i.e., if the results of $x_1=i$ and $x'_1=j$ are equal), and $1$ if $b_1\neq b'_1$. From assumptions (i)--(iii), it follows that $A_{ij}$ is sharp and compatible with $x_1=i$ (the measurement of the first party in Stockholm) and $x'_1=j$ (the measurement of the first party in Vienna), since $A_{ij}$ is realizable by joining outcomes $(0,0)$ and $(1,1)$ of the sharp measurements $(i,j)$ into outcome $A_{ij}=0$, and outcomes $(0,1)$ and $(1,0)$ into outcome $A_{ij}=1$.

Now we recall assumption (iv): There exist {\em some} measurements $x_1=0$, $x_1=1$, $x'_1=0$, and $x'_1=1$ such that $A_{00}$ and $A_{11}$ are compatible and $A_{01}$ and $A_{10}$ are compatible. Notice that this assumption is weaker than assuming similar compatibility relations for those measurements that maximize $\Sigma_n$.

For these $x_1=0$, $x_1=1$, $x'_1=0$, and $x'_1=1$, we can add a new event of the type either $(c,d|A_{00},A_{11})$ or $(e,f|A_{01},A_{10})$ to each of the sets of pairwise exclusive events constructed before in such a way that the new set contains only pairwise exclusive events. For example, to each of the sets of pairwise exclusive events for $n=2$ parties described in the left-hand side of Table~\ref{Table2} we can add the event displayed in the right-hand side of Table~\ref{Table2}.

For these $x_1=0$, $x_1=1$, $x'_1=0$, and $x'_1=1$, since each $A_{ij}$ can only give outcomes $0$ and $1$,
\begin{align}
\begin{split}
 &\sum_{c,d\in \{0,1\}} p(c,d|A_{00},A_{11})=1, \\
 &\sum_{e,f\in \{0,1\}} p(e,f|A_{01},A_{10})=1,
 \label{norm}
\end{split}
\end{align}
where $c, d, e, f$ are the outcomes of $A_{00}, A_{11}, A_{01}$, $A_{10}$, respectively.

Recall that assumptions (ii) and (iii) imply the E principle \cite{CY14}. Therefore, for each set of pairwise exclusive events, the sum of their probabilities cannot be larger than~1. This can be expressed as an inequality. We will refer to these inequalities as ``E inequalities.'' There is one E inequality for each of the $4^n$ sets of pairwise exclusive events constructed before. If we sum the corresponding $4^n$ E inequalities and take into account (\ref{norm}), we reach to the following inequality:
\begin{equation}
 \Sigma_n^2 + \left(2^n-\Sigma_n\right)^2 + 4^{n-1} \stackrel{\mbox{\tiny{E}}}{\leq} 4^n,
 \label{concl}
\end{equation}
where the origin of the terms $\Sigma_n^2$ and $\left(2^n-\Sigma_n\right)^2$ has been explained before, while $4^{n-1}$ follows from the fact that there are $4^{n-1}$ sums like those in Eq.~(\ref{norm}), and $4^n$ follows from the fact that we are summing $4^n$ E inequalities. See Supplemental Material \cite{SM} for extra details.

From inequality (\ref{concl}), we can conclude that, for those $x_1=0$, $x_1=1$, $x'_1=0$, and $x'_1=1$ such that $A_{00}$ and $A_{11}$ are compatible and $A_{01}$ and $A_{10}$ are compatible,
\begin{equation}
 \Sigma_n \stackrel{\mbox{\tiny{E}}}{\leq} (2 + \sqrt{2}) \times 2^{n-2}.
 \label{bound}
\end{equation}

The upper bound in inequality (\ref{bound}) also holds when $A_{00}$ and $A_{11}$ are incompatible and/or $A_{01}$ and $A_{10}$ are incompatible. This follows from the following result.

{\em Theorem 1.---}For any GPT only restricted by assumptions (i)--(iv), if these assumptions enforce that
\begin{equation}
\label{doce}
 \sum_{e_i \in V} p(e_i) \le \vartheta(G)
\end{equation}
for a particular set of events $\{e_i\}$ whose exclusivity graph $G$ is vertex transitive, has vertex set $V$, and Lov\'asz number $\vartheta(G)$,
then
\begin{equation}
\label{trece}
\sum_{e'_i \in V} p(e'_i) \le \vartheta(G)
\end{equation}
for any other set of events $\{e'_i\}$ whose exclusivity graph is $G$.

We remind the reader that $G$ is vertex transitive if for any two vertices $\{v_1, v_2\} \in V$ there is some automorphism $f$ such that $f(v_1)=v_2$. $\vartheta(G)$ is defined in, e.g., Refs.~\cite{CSW10,CSW14,GLS88}.

{\em Proof.---}By contradiction. Let us assume that there is a set of events $\{e'_i\}$ such that $\sum_{e'_i \in V} p(e'_i) = \vartheta(G) + \epsilon$, with $\epsilon > 0$. Let us denote by $\overline{G}$ the complement of $G$ (i.e., the graph with the same vertex set as $G$ and such that two vertices are adjacent if and only if they are not adjacent in $G$).
Assumptions (i)--(iv) allow for the existence of a set of events $\{e''_i\}$ with exclusivity graph $\overline{G}$ and probability assignments satisfying $\sum_{e''_i \in V} p(e''_i) = \vartheta(\overline{G})$, for any $\overline{G}$. For instance, QT is an example of a GPT that satisfies both assumptions (i)--(iv) and that there is a set of events $\{e''_i\}$ with exclusivity graph $\overline{G}$ such that $\sum_{e''_i \in V} p(e''_i) = \vartheta(\overline{G})$, for any $\overline{G}$ \cite{CSW10,CSW14}.

Now suppose that $\{e'_i\}$ and $\{e''_i\}$ correspond to two independent experiments. If $G$ is vertex transitive, then the E principle enforces
\begin{equation}
 \left[\sum_{e'_i \in V} p(e'_i)\right] \left[\sum_{e''_i \in V} p(e''_i)\right] \stackrel{\mbox{\tiny{E}}}{\leq} |V|,
 \label{prod}
\end{equation}
where $|V|$ is the cardinal of $V$ \cite{ATC14} (see also the Supplemental Material \cite{SM}). However, if $G$ is vertex transitive, $\vartheta(G) \vartheta(\overline{G}) = |V|$ \cite{GLS88}, thus inequality (\ref{prod}) is incompatible with our initial assumption. \hfill \endproof

In our case, $\{e_i\}$ is the set of events in $\Sigma_n$. Their exclusivity graph is vertex transitive and has as Lov\'asz number the right-hand side of inequality (\ref{bound}). Therefore, Theorem 1 guarantees that inequality (\ref{bound}) also holds for any choice of local measurements. This finishes the proof, since the bound in inequality (\ref{bound}) is exactly the quantum bound in expression (\ref{newineq}).


{\em Conclusion.---}By adopting the framework of GPTs (as in, e.g., Ref.~\cite{CY14}) and making assumptions (i)--(iv), we have presented a simple explanation of the quantum limits of the $n$-body nonlocal correlations \cite{Svetlichny87,CGPRS02,SS02} recently observed in several experiments \cite{LKR09,EMFLHYPBPSRGLWJR14,HSHMMVMNRJ14,BBSNHMGB13}. This result is important because it shows that these limits and those of single-body contextuality \cite{Cabello13,ATC14} admit a common explanation: the exclusivity principle, which can be derived from assumptions (ii) and (iii). This unified explanation is specially important since it shows that all these limits for correlations naturally derive from physical principles independently proposed to reconstruct quantum theory \cite{CY14}. This points out the convergence between the two main programs towards understanding quantum theory, namely, identifying the principles of correlations and reconstructing the theory from physical postulates.




\begin{acknowledgments}
The author thanks A.\ Ac\'{\i}n, M.\ Kleinmann, A.\,J.\ L\'opez-Tarrida, J.-\AA.\ Larsson, G.\ Murta, M.\ Paw{\l}owski, M.\ Terra Cunha, and E.\ Wolfe for useful conversations and comments on this manuscript. This work was supported by Project No.\ FIS2011-29400 (MINECO, Spain) with FEDER funds and the FQXi large grant project ``The Nature of Information in Sequential Quantum Measurements.''
\end{acknowledgments}



\end{document}